\documentstyle[11pt]{article}
\begin{document}
\begin{center}
{\Large \bf Pattern recognition in high multiplicity events}       \\
\bigskip
{\large N.M. Astafyeva$^1$, I.M. Dremin$^2$, K.A. Kotelnikov$^2$}\\
\bigskip
{\normalsize $^1$Space Research Institute, 117810 Moscow, Russia\\
$^2$Lebedev Physical Institute, 117924 Moscow, Russia}
\end{center}
\begin{abstract}
In very high energy collisions, many particles are produced and distributed in
the available phase space volume in various ways. With advent of new accelerator
facilities (especially, for nucleus-nucleus collisions), the problem of
pattern recognition in high multiplicity events becomes very actual in order to
classify such events and to separate those of them with some special features
(e.g. ring-like events, jets and other regular patterns).
This paper presents the first attempt to use
wavelet technique for pattern recognition in nucleus-nucleus collisions. After
describing the method in general, we demonstrate its power by applying it to
a single event of Pb-Pb collision at 158A GeV with 1072 charged particles
produced and discuss results obtained for some other events. Further extension
of the method is proposed.
\end{abstract}
High energy collisions of particles and nuclei result in production of many
secondary particles. Events with numbers of charged particles produced exceeding
several hundreds or a thousand are not the rare guests in cosmic ray studies
or at colliders. Moreover, the experiment ALICE at the newly built accelerator
LHC is planning to register events with about 10000 final particles created in
nucleus-nucleus collisions. Similar numbers are typical for high energy cosmic
ray showers registered usually by X-ray film technique. In recent emulsion
experiments in CERN some events with more than 1000 charged tracks were detected
in Pb-Pb collisions at 158A GeV.

Apart the common difficulties with particle registration, identification and
measuring all their characteristics, the further problem of pattern recognition
appears when considering particle distributions within the available phase
space. Let us stress here that such problems are of interest both for 
physicists and for computing community.
Traditional methods of one-dimensional projections on some axis (e.g.,
rapidity distribution etc) are very helpful but conceal many intriguing
patterns, sometimes directly seen by eye. For example, the behavior of such
distributions at ever smaller rapidity bins reveals its intermittent (fractal)
origin (for the review, see \cite{1}) but shows it in the integral form only.
The two- or three-dimensional pattern
could be, however, much reacher in structure compared to its one-dimensional
projection. In particular, Lego-plots have been used for jets, and
some ring-like densely populated regions
(corresponding to semi-isolated dense groups on the pseudorapidity axis) have
been observed in experiment \cite{a,n}.

We demonstrate the similar patterns for a single event with 1072 charged
tracks and propose the effective method of pattern recognition in high
 multiplicity events
based on the wavelet analysis. For brevity purposes, we consider just a single
event to clearly show the power of the method leaving high statistics analysis
for more detailed subsequent publications.

First, we describe the traditional method of analysis and its results. Then,
the brief introduction to wavelets is given and the method is applied to the
event analysis. Physics implication of its results and their comparison  to
traditional ones are discussed at the end together with some comments on other
events considered and future prospects for high statistics samples.

As an example to be demonstrated and treated in the paper, we have chosen the
event of Pb-Pb interaction at energy 158A GeV with 1072 charged tracks,
registered by EMU-15 Collaboration \cite{2}. Its target diagram (i.e. traces of
charged particles left in the plane perpendicular to the collision axis) is
shown in Fig.1. The collision axis passes through the center of the diagram
perpendicular to the plane so that the circle surrounding it corresponds to
different azymuthal angles $\phi$ at a fixed polar angle $\theta $ measured
by its radius. Two such circles are drawn for pseudorapidities $\eta =
-\log \tan \frac {\theta }{2}$ equal to 2.6 and 3.0 to demonstrate the scale.

By eye, one can easily recognize the regions densely and weakly populated by
particles. Being projected onto the pseudorapidity axis, the event provides
the histogram shown in Fig.2. Its structure looks less rich than that of the
two-dimensional plot but still there are maxima and minima, especially in the
region $1.7<\eta <3$, which could raise some speculation on their physics
origin. In particular, in previous publications on the subject \cite{2,3}
they were considered as an indication on the ring-like substructure inherent to
this event. Here, we leave aside any physics interpretation of this 
substructure having in mind, however, several possible candidates like
Cherenkov gluons, disoriented chiral condensate etc for further study.
We show that wavelet analysis gives more complete and quantitative
results. The idea behind this method is to resolve any pattern at different
locations with variable resolution. By studying the multiparticle final state
within various scales one can learn about space-time evolution of the process
as a whole (recall the inside-outside cascade, jet evolution etc).

Let us remind (for recent review see \cite{4}) that the wavelet transform of any
function of a single variable $f(x)$ is written as
\begin{equation}
W_{f}(a,b)=\vert a\vert ^{-1/2} \int f(x)\psi (\frac{x-b}{a})dx ,  \label{1}
\end{equation}
where the function $\psi (x)$ is called a wavelet. In what follows, we use
MHAT (Mexican hat) wavelet
\begin{equation}
\psi (x)=(1-x^2)\exp (-\frac {x^2}{2}).   \label{2}
\end{equation}
Thus one gets the two-parameter representation of the one-variable function.
From formulae (1) and (2) one sees that the function $f(x)$ is analysed close
to $x=b$ with a "spatial" scale resolution defined by the effective
"window width" $a$.
In particular, it resolves any pattern at a given point within different scales
("a mathematical microscope"). That is why it can be applied to the problem
under consideration.

The wavelet transform squared is conventionally called "the power spectrum"
$E_{W}$ in complete analogy to its Fourier transform analogue because it is
related to the squared function $f(x)$ by the following formula
\begin{equation}
\int f^{2}(x)dx=C_{\psi }^{-1}\int \int W^{2}_{f}(a,b)\frac {dadb}{a^2}, \label{3}
\end{equation}
where $C_{\psi }$ is a constant depending on the wavelet chosen. Therefore,
$E_{W}=W^2$ describes the "density" of the analyzed signal $f(x)$ in the
two-dimensional space $(a,b)$, i.e. its $a$-scale component at the location $b$.
When plotting equal-height levels of $E_W$ on the plane $(a,b)$, one resolves
the underlying pattern given by the function $f(x)$.

To exemplify analytically wavelet transforms, we show their general structure
for a very simple harmonic function $f(x)=\sin \omega x$. It is easy to get
\begin{equation}
E_W \propto a^{-3}(a\omega )^{4}\exp [-a^{2}\omega ^{2}]\sin ^{2}b\omega . \label{x}
\end{equation}
For constant $\omega $ and $a$, it is periodic along $b$-axis $E_W \propto
\sin ^{2}b\omega $ with a period inverse to $\omega $. For constant $\omega $
and $b$, it has a maximum at $a=(\sqrt{2} \, \omega )^{-1}$ and decreases
fast at larger values of $a$. Its intensity in the maximum increases as
$\omega ^{3}$ with $\omega $ increasing. For varying $\omega $, it possesses
faster oscillations along the $b$-axis and faster decrease in the $a$-direction
at higher $\omega $. Thus, looking at the extention of the power spectrum in
$b$ and $a$, one learns about the $\omega $-components at the stage.

We have applied this method to the event described above. To do that, we split
the plane $\phi ,\eta $ of the target diagram Fig.1 into 24 equal sectors with
$\Delta \phi =\pi /12$. In each sector labeled by $1\leq j\leq 24$ the function
\begin{equation}
f_{j}(\eta )=\sum _{i=1}^{n_{j}}\delta (\eta - \eta_{i})   \label{4}
\end{equation}
describing the particle density is obtained. Here $\delta (\eta - \eta_{i})$
is the Dirac delta-function, $n _{j}$
denotes the number of particles within the sector $j$, $\eta _{i}$ is the $\eta$
coordinate of $i$-th particle in that sector.

All 24 functions $f_j$ have been analyzed according to eq.(1) and their power
 spectra $E_W$ have been plotted by computer. They are represented by various
 squared sums of wavelets (2) located at $\eta _i$. In Fig.3 from top to
bottom, we show the
equal-height levels (denoted by different density black regions) of the 
corresponding spectra for sectors 3, 5, 7 and 21 (i.e., $(k-1)\pi /12<\phi _k<
k\pi /12$). Their most remarkable common feature is the dark strips indicated
by arrows which appear in the most densely populated regions on the
$(\phi ,\eta )$ plane. They are rather narrow in $\eta $ (as seen from their
extension along the $b$-axis) and cover an extended part of all available 
azymuthal angles $\phi $ (since they propagate to many $\phi $-sectors). 

It reminds the part of a ring (or an ellipse) rather dense with particles in
the upper, lower and right-hand sides and somewhat diluted to the left. The ring
is not centered, however, around the collision axis, and its center is located
at $\eta _{0} \approx 2.5,\phi _{0} \approx -\pi /4$, i.e. somewhat aside 
this axis. From positions of the arrows in Fig.3 one can notice that
the azymuth-asymmetric ring covers those regions of pseudorapidity $\eta $
where the peaks (which provoked the conclusion \cite{2,3} on ring-like structure)
are located on the pseudorapidity plot of Fig.2. However due to its non-centered
position it is not so easy to guess this new structure from the traditional
analysis of Fig.2 directly since it is smoothed there while the wavelet 
analysis reveals it and provides the more detailed information about it.

Another feature of the event shown by such an analysis is the existence of several
spot-like regions of various scales (sometimes with short fractal substructure).
They are seen as shorter dark regions (with tree-like branching in case of
fractals). They would correspond to the densely populated regions of small
extension in both polar and azymuthal angles i.e. to narrow jets which are
usually detected as towers on Lego-plots. Here they are not
however as typical as for some other events studied by us. Nevertheless, we
show in Fig.4 such strong jet-like structure seen just in the sector 23 and not
extending to the neighbouring $\phi $-sectors. The number of dark strips at 
the very small values of $a$ as well as the darkness of the spot indicate the
multiplicity within such a mini-jet while its $a$-extension shows its 
pseudorapidity width (see the formula (4) for a hint).

Let us note that the wavelets are always the functions
with both positive and negative values because their integral should vanish
(see review \cite{4}). Therefore, the power spectrum of a single isolated
particle contains three strips in the case of the Mexican wavelet as seen in
the right-hand side of Fig.3. The sideway strips are 
directed along $\eta =\eta _i \pm \sqrt {3} a$ and are about five times less
intensive than the vertical one located at $\eta =\eta _i $.
 Still looking in the Figs., one should
remember that the minima could become dark as well if the particles come
closer. It does not happen if the wavelet coefficients (not their squared
values!) are plotted with negative values cut off. It can be used for
detailed interpretation of each strip to be done in a subsequent
publication.

We postpone also the discussion of larger
statistics samples for a more complete publication. We would just mention that
the power spectrum wavelet analysis of the studied events reveals a variety
of patterns with some typical and self-similar features. It provides a hope
for the unique classification of high multiplicity events.
Let us say about some of them. Sometimes the whole event looks very "spotty" with
different (and sometimes "quantized") spot sizes that can be interpreted as
multiple mini-jets or clusters with some preferred 
values of multiplicity for a mini-jet.
Also the holes near the collision axis in Fig.1 (see the empty region in the
left-hand side of Fig.3) should be more carefully looked at. They imply some
peculiarities in the projectile fragmentation region. 
The target fragmentation region (the
right-hand side of Fig.3) has remarkably different structure in the wavelet
plot for different events as well. Their study is important for 
fragments escaping in beam pipes. The two-dimensional pattern of events
varies strongly reminding various geometrical figures (ellipses, "flowers"
etc). The more accurate two-dimensional wavelet analysis is in progress.

To conclude, we have shown in the example of a single high multiplicity event
that the wavelet analysis helps reveal event pattern in the available phase
space providing quantitative measures for its typical features in terms of
the location $b$ and the scale $a$. Moreover we have found that the wavelet
plots of different events differ but can be grouped in such a way to provide
the basis for their classification. This topic requires however more extended
publication which is being prepared.

\newpage
ACKNOWLEDGEMENTS

We are grateful to all members of EMU-15 Collaboration for the permission to
use their data for the analysis and, especially, to A.G. Martynov for his
invaluable help with Figures.
This work was supported in part by the Russian Fund
for Basic Research grants 96-02-19572 and 96-01-00340.

\begin{center}
FIGURE CAPTIONS
\end{center}
\begin{tabbing}
Fig.1 \= The target diagram of EMU-15 event of Pb-Pb collision at 158A GeV \\
      \> with 1072 charged tracks. Two circular lines are shown for $\eta =2.6$ \\
      \> and $\eta =3.0$. \\
Fig.2 \= The pseudorapidity distribution of particles in the event shown in Fig.1.\\
      \> The peaks in the region $1.7<\eta <3$ are related to the suspected \\
      \> ring-like structure.  \\
Fig.3 \= The power spectrum plots of the wavelet analysis for the event in Fig.1.\\
      \> The most dark regions correspond to the densely populated parts. The\\
      \> pseudorapidity increases along the abscissa axis, the scale $a$ increases\\
      \> down the ordinate axis. The Figures from top to bottom are obtained for\\ 
      \> azymuth sectors numbered by $j=3,5,7,21$ correspondingly (see the text). \\
      \> Arrows indicate the location of the ring-like structure in that sector.           \\
Fig.4 \= The jet-like pattern revealed by the wavelet analysis in the sector 23.\\
      \> The strong dark spot (shown by the arrow) does not extend very far\\
      \> to the neighbouring sectors.\\
\end{tabbing}

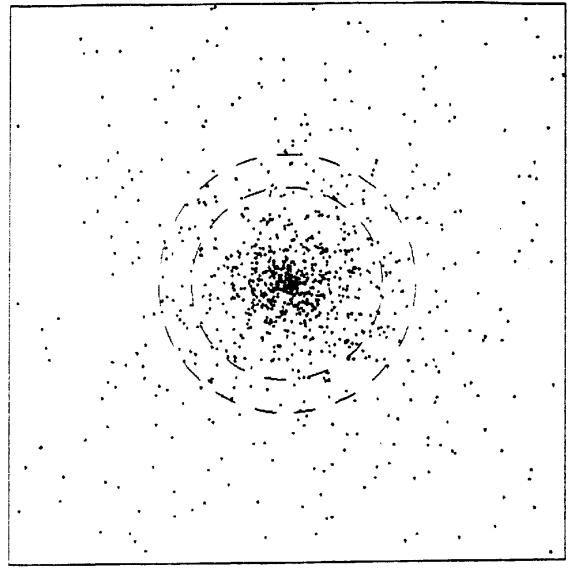
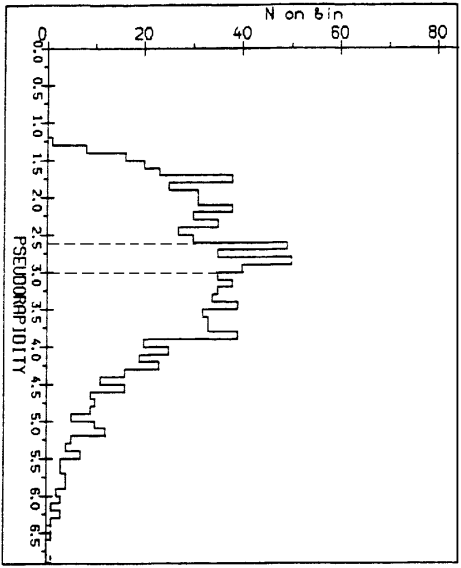
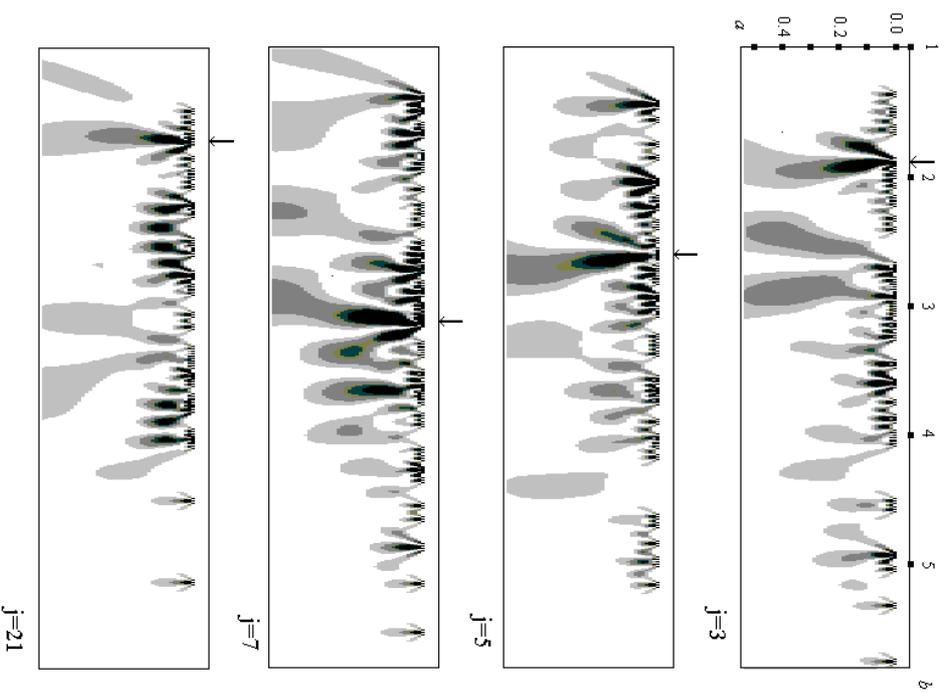
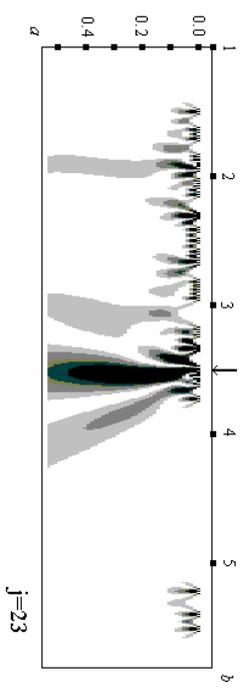
\end{document}